# Polarization properties of a symmetrical and asymmetrical nonreciprocal chiral photonic bandgap structure with defect

Vladimir R. Tuz[*]

*Department of Theoretical Radio Physics, Kharkov National University, Svobody Square 4, Ukraine*

[*]*Corresponding author: tvr@vega.kharkov.ua, Vladimir.R.Tuz@univer.kharkov.ua*

The polarization properties of perfectly periodical and defective one-dimensional photonic bandgap structures with nonreciprocal chiral (bi-isotropic) layers are studied. The method of solution is based on the $2\times 2$-block-representation transfer-matrix formulation. Numerical simulations are carried out for different type of structures (symmetrical or asymmetrical) in order to reveal the dependence of the reflection and transmission coefficients on frequency, chirality, nonreciprocity parameters, and the angle of wave incidence.

## 1. INTRODUCTION

The research on artificial chiral media has been especially intense within microwaves in the past two decades [1-4], and chirality is an example of novel material effects at best. In general, the nonreciprocal chiral (bi-isotropic) media are defined via the following constitutive equations where the electric and the magnetic inductions are related to the electromagnetic field intensity via the permittivity, permeability and magnetoelectric interaction parameters (the gyration parameters in terms of crystalloptics):

$$\mathbf{D} = \varepsilon\mathbf{E} + \xi\mathbf{H}, \quad \mathbf{B} = \mu\mathbf{H} + \zeta\mathbf{E}. \qquad (1)$$

Here $\xi = \chi + i\rho$ and $\zeta = \chi - i\rho$; $\chi$ describes the degree of inherent nonreciprocity in the medium (Tellegen parameter), $\rho$ is a measure for handedness of the material (chirality parameter). As known, the eigenwaves in homogeneous chiral media are two circularly polarized (right- and left-handed) waves which exhibit some interesting behaviors [4]. Thus, the real part of the chirality parameter $\rho$ defines the optical rotatory dispersion, which causes a rotation of polarization due to different phase velocities of the right- and left-handed circularly polarized wave. Then, the imaginary part of $\rho$ and Tellegen parameter $\chi$ define the circular dichroism, which modifies the nature of polarization of the propagating wave by causing an ellipticity (due to a difference between the absorption coefficients of the right- and left-handed circularly polarized waves and nonorthogonality of electric and magnetic field vectors). The effect of the optical activity in chiral media can by so strong, that the refractive index becomes negative, even if the permittivity $\varepsilon$ and permeability $\mu$ are both positive. Specifically, this negative refraction may occur for one circularly polarized wave (backward wave), while for the other circular polarization the refractive index remains positive. These properties have attracted considerable attention to chiral media and may open new potential applications in optics [5-8].

In the present paper, we consider the chiral media in their application to constructing photonic bandgap crystals [9-11]. There are a lot of publications where both ideally periodic and quasi-periodic [12-14] photonic crystals are studied, including the group of chiral photonic crystals [15-19]. In view of the mentioned above behaviors of chiral media, the chiral photonic crystals possess rich optical properties and are characterized by the selective reflection in a specific waveband in addition to circular polarization of the reflected and transmitted light. Such artificial structures are widely used in the modern integrated optics and optoelectronics, laser techniques, millimeter and submillimeter wave devices, and optical communications.

Nonreciprocal photonic structures are also of special interest. The nonreciprocity is a comparable effect to the chirality from theoretical point of view, but in terms of manufacturing real-life samples, nonreciprocal materials are not as advanced [3]. It is known that, externally, the nonreciprocity can be created by a static magnetic field in ferrites or plasmas, but in those cases the medium is anisotropic. In recent years, some new mechanisms of obtaining nonreciprocity have been proposed [20, 21] in addition to natural nonreciprocal materials [22, 23]. In particular, the importance of nonreciprocal photonic crystals is related, to possibility of creating omnidirectional reflectors, optical diodes, and etc. [24-26]. The optical diode is a device transmitting light in one direction and blocking it in the reverse direction (i.e., characterized by a nonreciprocal transmission). Various types of optical diodes have been proposed. The well-known optical insulator based on linear polarizers and Faraday rotator is characterized by a low level of losses and a high ratio of the direct transmission to the reverse attenuation. Optical diodes can also be based on combinations of some other effects [7, 27].

The present paper is devoted to study of both one-dimensional perfectly periodic chiral photonic structures and those containing a defect, with taking into account the media nonreciprocity. We also consider an asymmetric defective structure which has two multilayer sections with the mirror handedness of chirality and nonreciprocity parameters on either sides of the defective element for the purpose of the optical diodes constructing. The method of solution is based on the $2 \times 2$-block-representation transfer-matrix formulation proposed earlier in [18].

## 2. PERFECTLY PERIODIC NONRECIPROCAL CHIRAL PHOTONIC STRUCTURE

Let's first consider a finite one-dimensional periodic structure of $N$ basic elements (periods) (Fig. 1). Each of periods includes a homogeneous isotropic (with permittivities $\varepsilon_1$, $\mu_1$) and bi-isotropic (with $\varepsilon_2$, $\mu_2$, $\xi$, $\zeta$) layers with thicknesses $d_1$ and $d_2$, respectively. The total length of structure period is $L = d_1 + d_2$. In general, the material parameters $\varepsilon_1$, $\mu_1$, $\varepsilon_2$, $\mu_2$, $\chi$, $\rho$ are frequency dependent and complex for lossy media. In a particular case of $\rho \neq 0$, $\chi = 0$ the structure layers are chiral and reciprocal (the Pasteur layers), when $\rho = 0$, $\chi \neq 0$ they are achiral and nonreciprocal (the Tellegen layers). The outer half-spaces $z \leq 0$ and $z \geq NL$ are homogeneous, isotropic and have permittivities $\varepsilon_0$, $\mu_0$ and $\varepsilon_3$, $\mu_3$, respectively.

As the excitation fields, the plane monochromatic waves $\left[\exp(-i\omega t)\right]$ with perpendicular ($\mathbf{E}^e \parallel \mathbf{x}_0$, $H_x^e = 0$, $s = e$) or parallel ($\mathbf{H}^h \parallel \mathbf{x}_0$, $E_x^h = 0$, $s = h$) polarization are selected. They are obliquely incident from the region $z \leq 0$ at an angle $\varphi_0$ to the $z$ axis.

Making use of the transfer-matrix formalism [28], the equation coupling the field amplitudes at the structure input and output for the incident fields of $E$ type ($s = e$) and $H$ type ($s = h$) is obtained in [18] as

$$\mathbf{V}_0 = \mathbf{T}\mathbf{V}_{N+1} = \left(\mathbf{T}_{01}\mathbf{T}^{N-1}\tilde{\mathbf{T}}\right)\mathbf{V}_{N+1} = \left(\mathbf{T}_{01}\mathbf{T}^N\mathbf{T}_{13}\right)\mathbf{V}_{N+1}, \qquad (2)$$

where $\mathbf{V}_0 = \left\{A_0^s \ \ B_0^s \ \ 0 \ \ B_0^{s'}\right\}^T$ and $\mathbf{V}_{N+1} = \left\{A_{N+1}^s \ \ 0 \ \ A_{N+1}^{s'} \ \ 0\right\}^T$ are vectors containing the field amplitudes at the structure input and output, $T$ is the matrix transpose operator, $\mathbf{T}_{01}$, $\mathbf{T}$, $\tilde{\mathbf{T}}$ are the transfer-matrices of the illuminated boundary, the repeated heterogeneity, and the last element, which is loaded on the waveguide channel having the admittance $Y_3^s$. We denote here $A^s$, $A^{s'}$ and $B^s$, $B^{s'}$ as the amplitudes of co-polarized ($s$) and cross-polarized ($s'$) components of the transmitted and reflected fields, respectively. The elements of the transfer-matrices (2) are determined from solving the boundary-value problem and are presented in [18, 29].

The algorithm from the matrix polynomial theory [30] for raising the matrix $\mathbf{T}$ to the power $N$ was introduced in [31] to study the structure with a large number of periods ($N \gg 1$)

$$\mathbf{V}_0 = \left[ \mathbf{T}_{01} \left( \sum_{n=1}^{4} \lambda_n^N \mathbf{F}_n \right) \mathbf{T}_{13} \right] \mathbf{V}_{N+1}. \qquad (3)$$

Here $\lambda_n$ are the eigenvalues of the transfer-matrix $\mathbf{T}$, $\mathbf{F}_n = \mathbf{P}\mathbf{I}_n\mathbf{P}^{-1}$, $\mathbf{P}$ is the matrix which columns are the set of independent eigenvectors of $\mathbf{T}$, $\mathbf{I}_n$ is the matrix with a 1 in the ($n$, $n$) location and zeros elsewhere.

From (3), the required reflection and transmission coefficients are determined by the expressions $R^{ss} = B_0^s / A_0^s$, $\tau^{ss} = A_{N+1}^s / A_0^s$ and $R^{ss'} = B_0^{s'} / A_0^s$, $\tau^{ss'} = A_{N+1}^{s'} / A_0^s$ for the co-polarized and cross-polarized waves, respectively.

The parametrical dependences of the reflection and transmission spectra of plane monochromatic waves of the finite structure with bi-isotropic layers have interleaved areas with high (the stopbands) and low (the passbands) average level of the reflection (Figs. 2 and 3). Due to an interference of the reflected waves from outside boundaries of layers $N-1$ small-scale oscillations appear in the passbands.

As known, the eigenwaves of an infinite convenient isotropic medium have transverse magnetic and transverse electric polarizations, or their linear combinations, and, in general, they have the elliptical polarization. On the other hand, in a bi-isotropic medium, fields can be split into left- and right-hand circularly polarized eigenwaves which have different propagation constants $\gamma^\pm$, and each of this eigenwaves sees the medium as if it were an isotropic medium with equivalent parameters $\varepsilon^{\pm}$ and $\mu^{\pm}$ [2, 18, 29]. Right- and left-hand circular polarizations are also the eigenwave polarizations in cholesterics. However, while the photonic band gap in cholesterics exists only for one circular polarization that coincides with the chirality sign of the medium, in chiral periodic media both circular polarizations are diffracting polarizations. From this characteristic of chiral periodic structures follows the independence of their reflection spectra from the chirality parameter $\rho$ in case of normal wave incidence; a chiral periodic structure and an achiral one with the same parameters are characterized by identical reflection and transmission spectra (the stopband and passband positions). The medium chirality causes only change the rotation of polarization plane of the transmitted wave. The theoretical angle of the polarization rotation in a chiral medium is given by $\alpha = -\text{Re}(\rho)k_0 z$ [2, 4], where $\text{Re}(\rho)$ is the real part of the chirality parameter, $k_0$ is the wavenumber in free space, and $z$ is the distance passed by the wave through chiral layers. The complexity of $\rho$ modifies the nature of the propagating wave by introducing ellipticity. This is due to the different absorption coefficients for the right- and left-handed circularly polarized waves. The theoretical ellipticity depends on the imaginary part of the chirality as follows [4]: $\eta = \left|\left(\exp[2k_0 z \text{Im}(\rho)] + 1\right) \big/ \left(\exp[2k_0 z \text{Im}(\rho)] - 1\right)\right|$, where $\eta$ is the ellipticity defined as the ratio between the major and minor axes of the ellipse, $\text{Im}(\rho)$ is the imaginary part of the chirality parameter. As follows from our numerical calculation, the cross-polarized components of reflected and transmitted waves are equal to each other $|R^{eh}| = |R^{he}|$, $|\tau^{eh}| = |\tau^{he}|$ for the structure with reciprocal chiral layers (the Pasteur layers). The nonzero value of the nonreciprocal (Tellegen) parameter ($\chi \neq 0$) causes $|R^{eh}| \neq |R^{he}|$, and the cross-polarized component appears in the reflected field even for normal incidence ($\varphi_0 = 0$) of the exciting wave. The additional effects by the media nonreciprocality are the transmission spectra symmetry breakdown related to $\rho = 0$ and a disturbance of the amplitude of small-scale oscillations in the passbands for both the co-polarized and cross-polarized waves (Figs. 2 and 3).

At oblique incidence of the exciting wave, the pattern of the structure reflection and transmission becomes more complicated. The rotation spectra acquire a pronounced diffraction character: the rotation and ellipticity are heavily suppressed in the photonic band gap and strongly oscillate near the band gap boundaries. Here, the angle of rotation increases in one wavelength range and decreases in another. At large angles of incidence, the cross-polarized components of the reflected and transmitted fields equal zero ($|R^{eh}|=|R^{he}|=|\tau^{eh}|=|\tau^{he}|=0$). Thus, there appears a possibility for the diffraction control of the polarization rotation.

## 3. SYMMETRICAL AND ASYMMETRICAL NONRECIPROCAL CHIRAL PHOTONIC STRUCTURE WITH DEFECT

In this section we consider a nonreciprocal chiral photonic structure with a periodicity defect which is created via replacing the *m*-th period of the sequence (Fig. 4). In general, material parameters of periods of both structure sections, coming before and after of the defective element, are different. Their difference is defined with some prime substitution above the parameter. Thus, for the first, the last structure sections and the defect element we have: $\{\varepsilon_2', \mu_2', \xi' = \chi' + i\rho',$ $\zeta' = \chi' - i\rho'\}$, $\{\varepsilon_2'', \mu_2'', \xi'', \zeta''\}$ and $\{\varepsilon_2''', \mu_2''', \xi''', \zeta'''\}$, respectively. Without loss of generality, the material parameters of isotropic slabs and the layer thicknesses are left unchanged, i.e. $\varepsilon_1' = \varepsilon_1'' = \varepsilon_1''' = \varepsilon_1$, $\mu_1' = \mu_1'' = \mu_1''' = \mu_1$ and $d_j' = d_j'' = d_j''' = d_j$, $j = 1, 2$. It is defined that the structure is symmetrical when $\operatorname{Re}(\rho^{/}) = \operatorname{Re}(\rho^{//})$, $\operatorname{Re}(\chi^{/}) = \operatorname{Re}(\chi^{//})$, and asymmetrical when $\operatorname{Re}(\rho^{/}) = -\operatorname{Re}(\rho^{//})$ or $\operatorname{Re}(\chi^{/}) = -\operatorname{Re}(\chi^{//})$.

Instead of (2), the equation coupling the field amplitudes at the defective structure input and output is obtained as

$$\mathbf{V}_0 = \left\{ \mathbf{T}_{01} \left( \mathbf{T}^{/} \right)^{m-1} \mathbf{T}^{///} \left( \mathbf{T}^{//} \right)^{N-m} \mathbf{T}_{13} \right\} \mathbf{V}_{N+1} = \left\{ \mathbf{T}_{01} \left( \sum_{n=1}^{4} (\lambda_n^{/})^{m-1} \mathbf{F}_n^{/} \right) \mathbf{T}^{///} \left( \sum_{n=1}^{4} (\lambda_n^{//})^{N-m} \mathbf{F}_n^{//} \right) \mathbf{T}_{13} \right\} \mathbf{V}_{N+1}, \quad (4)$$

where $\mathbf{T}^{/}$, $\mathbf{T}^{//}$ and $\mathbf{T}^{///}$ are the transfer-matrices of the period of the first and the last structure sections and the defective element, respectively. Obviously, due to the non-commutativeness the matrix product, the spectra of the reflected and transmitted fields depend on the defective element position within the structure.

It is known that the defects inside a layered isotropic one-dimensional sample produce additional resonance modes (defective modes) in the stopbands. Such defects are widely used to produce high-$Q$ laser cavities in vertical-cavity-surface-emitting lasers. In analogy to the isotropic periodic structures, a defect can be produced in a chiral structure by adding an isotropic layer in middle of a sample (see [31], and its bibliography). In this case, the defective modes also exhibit a number of important polarization-related features.

The curves of the reflection and transmission spectra of a symmetrical and asymmetrical bi-isotropic photonic structure with a defective isotropic layer are presented on Figs. 5 and 6, respectively. For clarity, we consider separately the spectrum properties of the structure of two types. The structure of the first type consists of achiral, nonreciprocal layers (the Tellegen layers) [Figs. 5 (a, b) and 6 (a, b)], and the second type are chiral, reciprocal ones (the Pasteur layers) [Figs. 5 (c, d) and 6 (c, d)]. As can be seen in figs. 5 and 6, in addition to the mentioned above defective modes in the stopbands, there are some changes in the magnitude of the high-frequency (small-scale) oscillations in the passbands for both types of structures. Those effects are determined by the composition of the eigenmodes of the structure sections placed before and after the defective element, and by additional eigenmodes that appear as a result of the wave polarization transformation [31]. Especially interesting behavior of the perturbation of the mentioned small-scale oscillations shows for the structure with the Tellegen layers. In this case, the frequency bands ($4.0 < k_0 L < 4.5, 9.5 < k_0 L < 10.0$) of the high reflection of the cross-polarized field component appear in the passbands. At the same time, the co-polarized component dominates in the transmitted fields, and eventually, in these bands, we have the cross-polarized reflection and the co-polarized transmission $|R^{ss}| \approx 0$, $|R^{ss'}| \neq 0$ and $|\tau^{ss}| \neq 0$, $|\tau^{ss'}| \approx 0$.

The spectrum properties of the structure with the Tellegen layers in the stopbands are practically the same as the properties of the convenient defective isotropic periodic structure. Note only that there are the defective modes in the stopbands of both co-polarized and cross-polarized components of the reflected and transmitted fields. The reflection (transmission) level of the cross-polarized defective modes is much less than the co-polarized ones for the symmetrical structure [Figs. 5 (a) and 5 (b)], and these cross-polarized modes are practically absent for the asymmetrical structure, i.e. $|\tau^{ss}| \approx 1$ [Figs. 6 (a) and 6 (b)].

The defective mode behavior is more interesting in the structure with the Pasteur layers. As can be seen in figs. 5 (c) and 5 (d), there is a possibility of obtaining different combinations of the polarization of the reflected and transmitted fields using the symmetrical structure. For example, for the chosen structure parameters, we have: $|R^{eh}| \neq 0$, $|R^{ee}| \approx 0$, $|\tau^{ee}| \gg |\tau^{eh}|$ at $k_0 L \approx 2.6$ and $|R^{ee}| \approx |R^{eh}|$, $|\tau^{ee}| \approx |\tau^{eh}|$ at $k_0 L \approx 11.6$. The particular properties of the asymmetrical structure is that the cross-polarized component of the transmitted field is practically absent in the passbands and stopbands but the significant polarization transformation occurs on the band gap boundaries. The first effect is determined by the mutual discharging of the polarization rotation which provides two structure sections placed before and after the defective element with equal material parameters and mirror handedness. The oblique incidence of the wave and the mentioned diffraction character of the rotation near the band gap boundaries explain the second feature.

The further functionality expansion of the structure under study is connected with using of an anisotropic material to construct the defect. As is well known, in anisotropic media, uniform plane waves can be decomposed in two orthogonal polarization states (linear or circular) that propagate with two different speeds. The two states develop a phase difference as they propagate, which alters the total polarization of the wave. In the subject being discussed, the anisotropic defect can provide the difference of the defective mode configuration for the co-polarized and cross-polarized waves. The other way to complicate the spectrum features is an introduction of the nonlinear defect. It is very important for the optical diode constructing, because, as usually, the amplitude-frequency characteristic of the diode is strongly nonlinear.

We also study a finite isotropic periodic structure with the bi-isotropic defect. As above, we consider the nonreciprocal and chiral defects separately. The curves of the reflection and transmission spectra of the both structures are presented in Figs. 7 (a, b) and 7 (c, d), respectively. It is seen that the low level cross-polarized component appears in the reflected and transmitted fields, except in the level of the transmission for the Pasteur defective layer. On the other hand, the significant polarization transformation appears within the defective modes both in reflection and transmission. Note the value of the defective layer permittivity is higher than those in the structure sections ($\varepsilon_2^{///} > \varepsilon_2^{/}, \varepsilon_2^{/} = \varepsilon_2^{//}$). This condition provides the shifting of the defective mode frequency from the right (left) boundary of the low-frequency (high-frequency) stopbands.

We do not discuss here the spectral properties of the defective structure with simultaneous presence of the nonreciprocity and chirality of layers due to the limited size of the publication. It is clear that the integrated spectra will contain all the mentioned features in their combinations.

## 4. CONCLUSION

The polarization properties of both perfectly periodical and defective one-dimensional photonic bandgap structures are studied. The structure period consists of nonreciprocal, chiral (bi-isotropic) and isotropic layers. The periodicity defect is created via replacing the $m$-th period of the sequence with some element whose optical properties are different from the others. This defective element can be both isotropic and bi-isotropic. Two types of the structure (symmetrical and asymmetrical) are considered. They differ in the material parameters of periods of structure sections coming before and after of the defective element. This difference is defined via the handedness of the material of layers.

The dependences of the reflection and transmission spectra on the frequency, nonreciprocity, chirality parameters, and the angle of wave incidence are obtained. It is shown that the introduction of the media nonreciprocity and chirality changes the pattern of the structure reflection and transmission and causes the availability the cross-polarized component in the reflected field for the normal wave incidence, the perturbation of the small-scale oscillations in the passbands, a change of the defective mode configuration.

The width of photonic band gaps, as well as their spectral position and the distance between them, depend on the parameters of the problem and, thus, can be controlled. Therefore, such systems can be used as controllable polarization filters and mirrors, optical diodes, polarization transformers, mode discriminators, multiplexers for circularly or elliptically polarized waves, and sources of circular (elliptical) polarization.

List of Figure Caption

V. R. Tuz, Polarization properties of a symmetrical and asymmetrical nonreciprocal chiral photonic bandgap structure with defect

**Figure 1 (color online).** Finite photonic bandgap structure of isotropic and bi-isotropic layers.

**Figure 2 (color online).** (a) Frequency and (b) angular dependences of the reflection and transmission spectra of a finite photonic structure of isotropic and bi-isotropic layers. $\varepsilon_j = \mu_j = 1$, $j \neq 2$, $\varepsilon_2 = 2$, $\mu_2 = 1$, $\rho = 0.1$, $\chi = 0.05$, $d_1/L = d_2/L = 0.5$, $N = 5$. (a) $\varphi_0 = 25^0$. (b) $k_0 L = 10$.

**Figure 3 (color online).** Transmission coefficient magnitude of (a) co-polarized and (b) cross-polarized waves as function of the frequency $k_0 L$ and the chirality parameter $\rho$ for a finite photonic structure of isotropic and bi-isotropic layers. $\varepsilon_j = \mu_j = 1$, $j \neq 2$, $\varepsilon_2 = 2$, $\mu_2 = 1$, $\chi = 0.05$, $d_1/L = d_2/L = 0.5$, $N = 5$.

**Figure 4 (color online).** Finite photonic bandgap structure of isotropic and bi-isotropic layers with periodicity defect.

**Figure 5 (color online).** (a, c) Reflection and (b, d) transmission spectra of a finite symmetrical bi-isotropic photonic structure of $N = 19$ periods with the isotropic defect in the middle ($m = 10$) of the structure. $\varepsilon_j = \mu_j = 1$, $j \neq 2$, $\varepsilon_2' = \varepsilon_2'' = 2$, $\varepsilon_2''' = 1$, $\mu_2' = \mu_2'' = \mu_2''' = 1$, $\rho''' = \chi''' = 0$, $\varphi_0 = 25^0$, $d_1/L = d_2/L = 0.5$. (a, b) The Tellegen layers. $\rho' = \rho'' = 0$, $\chi' = \chi'' = 0.1$. (c, d) The Pasteur layers. $\rho' = \rho'' = 0.1$, $\chi' = \chi'' = 0$.

**Figure 6 (color online).** (a, c) Reflection and (b, d) transmission spectra of a finite asymmetrical bi-isotropic photonic structure of $N = 19$ periods with the isotropic defect in the middle ($m = 10$) of the structure. $\varepsilon_j = \mu_j = 1$, $j \neq 2$, $\varepsilon_2' = \varepsilon_2'' = 2$, $\varepsilon_2''' = 1$, $\mu_2' = \mu_2'' = \mu_2''' = 1$, $\rho''' = \chi''' = 0$, $\varphi_0 = 25^0$, $d_1/L = d_2/L = 0.5$. (a, b) The Tellegen layers. $\rho' = \rho'' = 0$, $\chi' = 0.1$, $\chi'' = -0.1$. (c, d) The Pasteur layers. $\rho' = 0.1$, $\rho'' = -0.1$ $\chi' = \chi'' = 0$.

**Figure 7 (color online).** (a, c) Reflection and (b, d) transmission spectra of a finite isotropic photonic structure of $N = 19$ periods with the bi-isotropic defect in the middle ($m = 10$) of the structure. $\varepsilon_j = \mu_j = 1$, $j \neq 2$, $\varepsilon_2' = \varepsilon_2'' = 2$, $\varepsilon_2''' = 3$, $\mu_2' = \mu_2'' = \mu_2''' = 1$, $\rho' = \rho'' = \chi' = \chi'' = 0$, $\varphi_0 = 25^0$, $d_1/L = d_2/L = 0.5$. (a, b) The Tellegen defective layer. $\rho''' = 0$, $\chi''' = 0.1$. (c, d) The Pasteur defective layer. $\rho''' = 0.1$, $\chi''' = 0$.

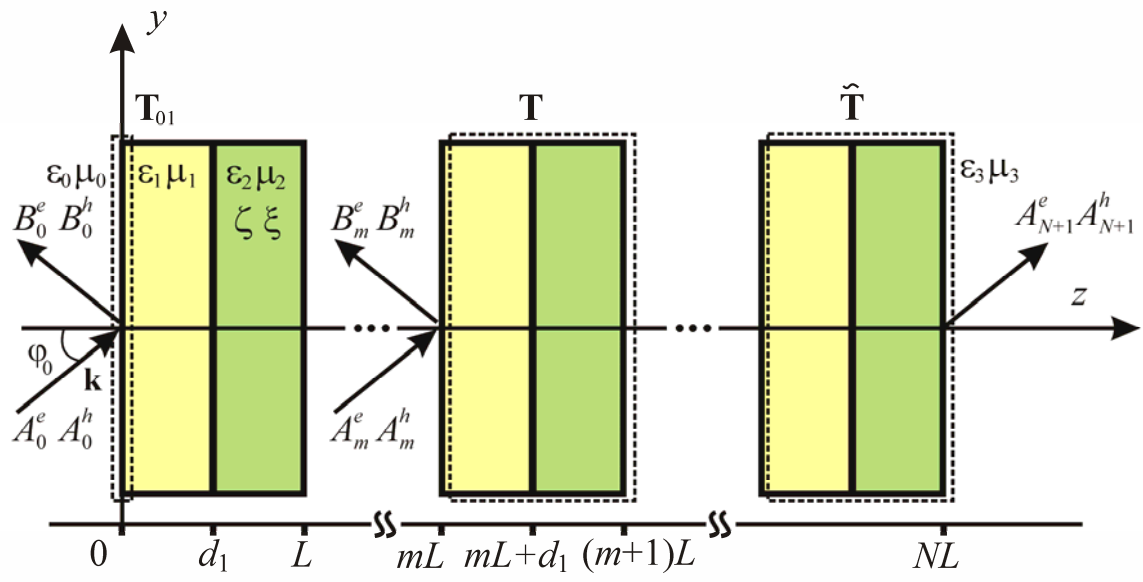

Figure 1

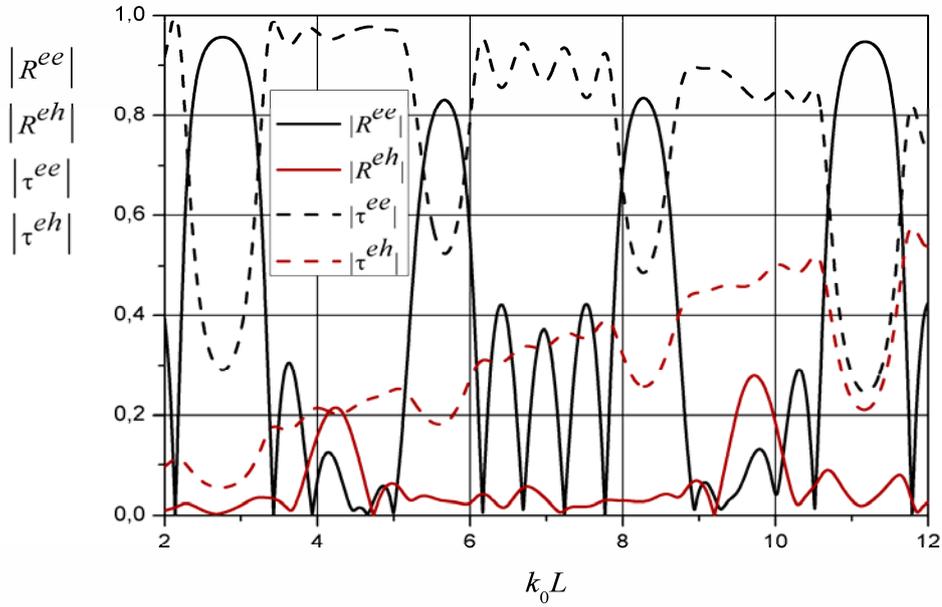

(a)

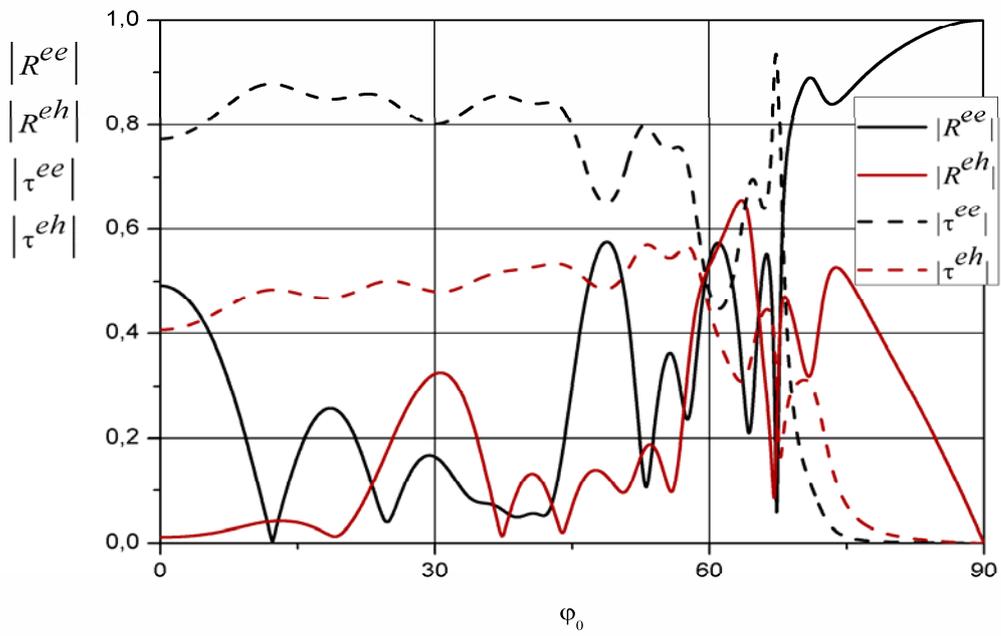

(b)

Figure 2

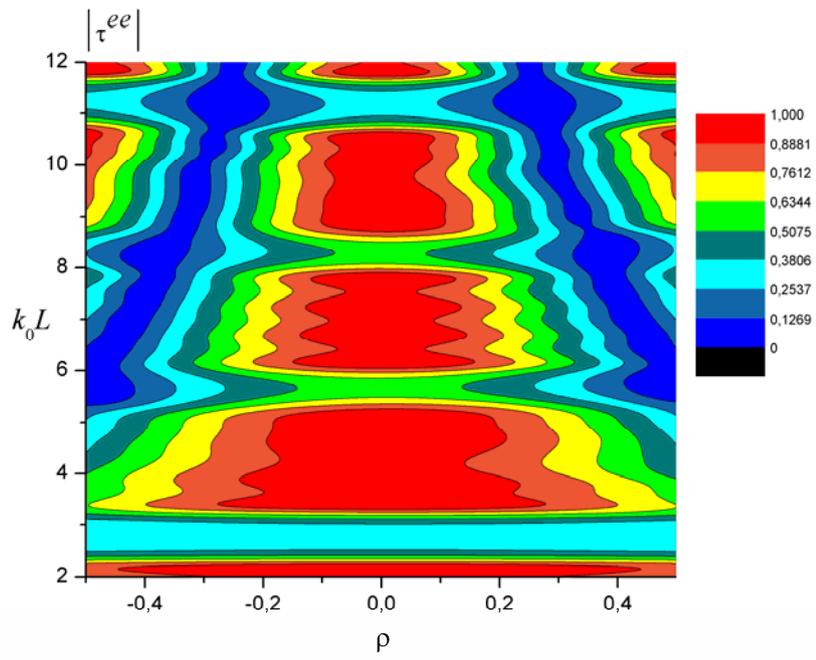

(a)

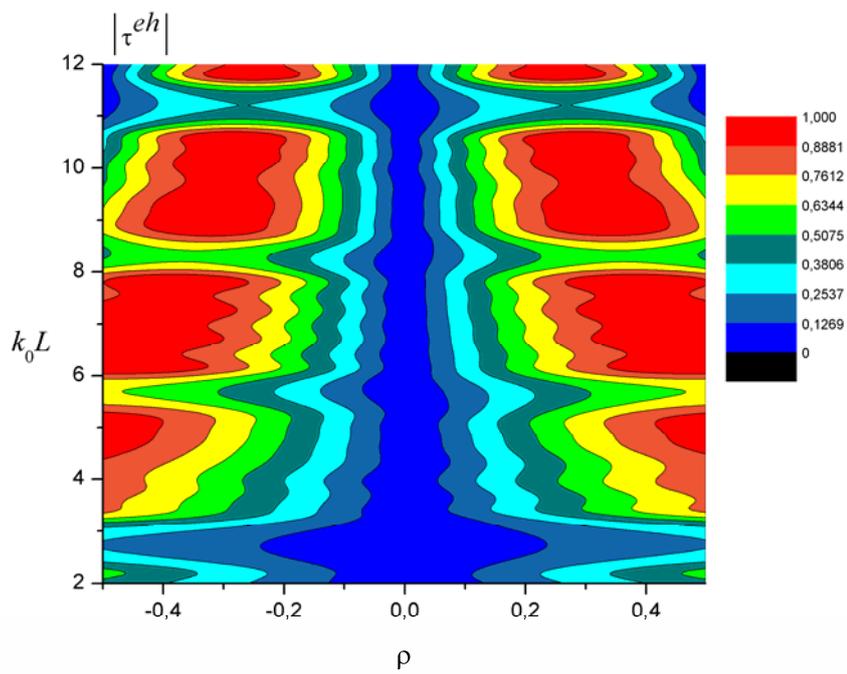

(b)

Figure 3

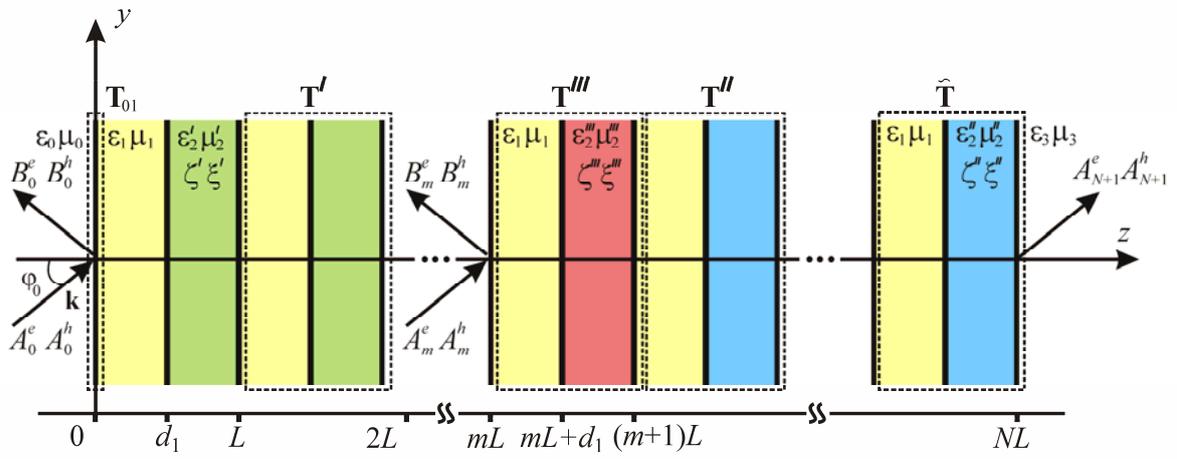

Figure 4

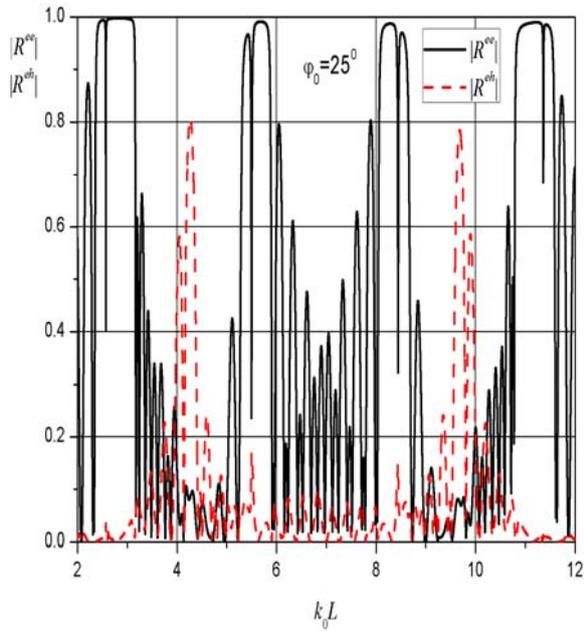

(a)

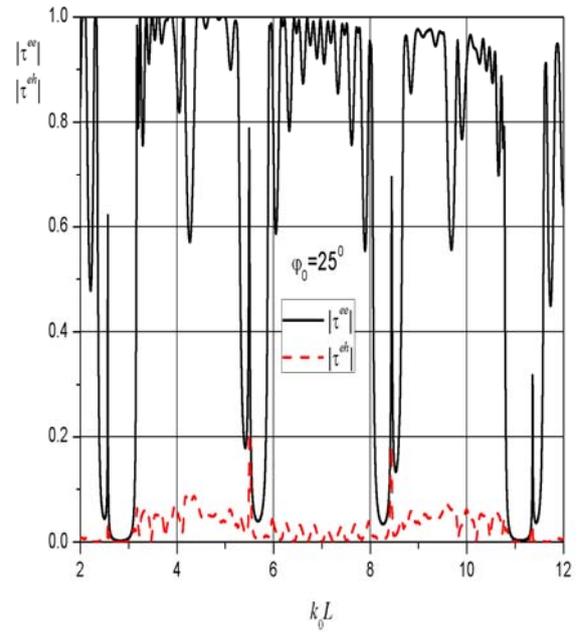

(b)

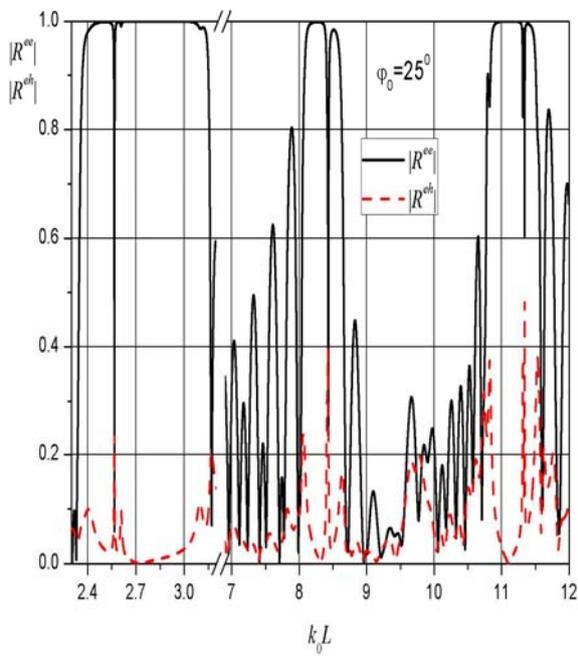

(c)

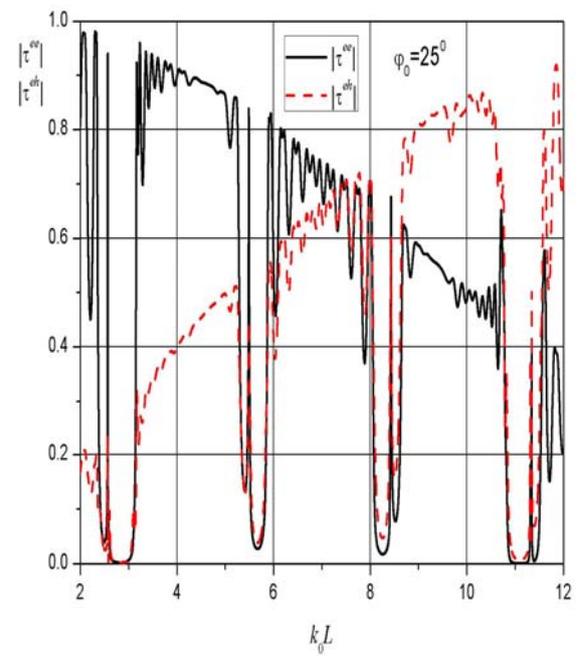

(d)

Figure 5

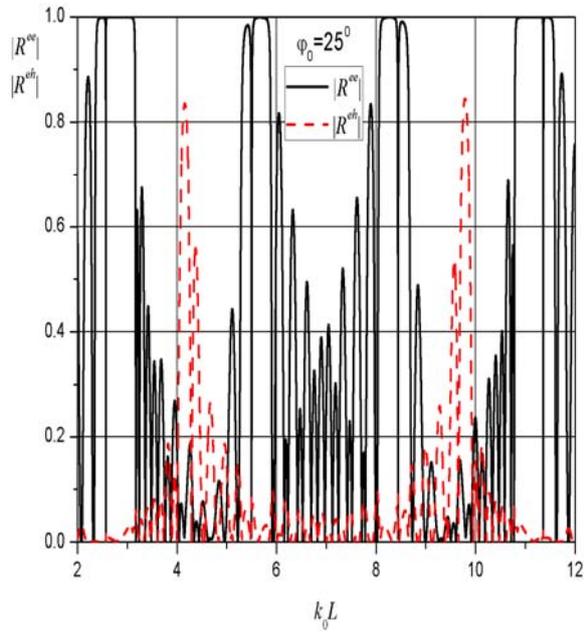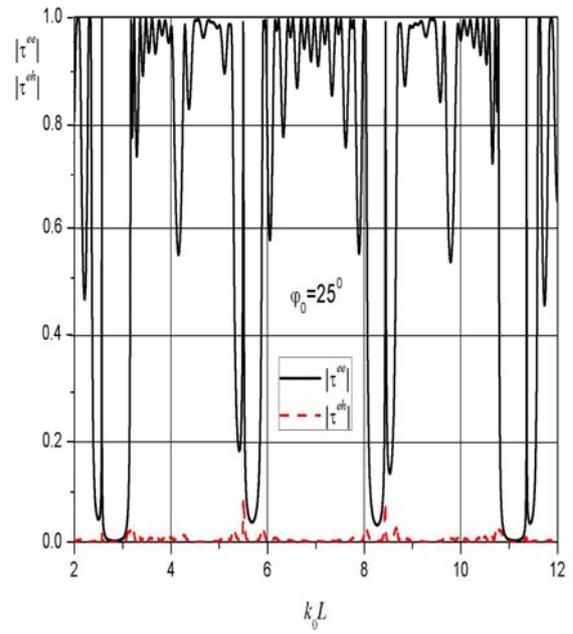

(a) (b)

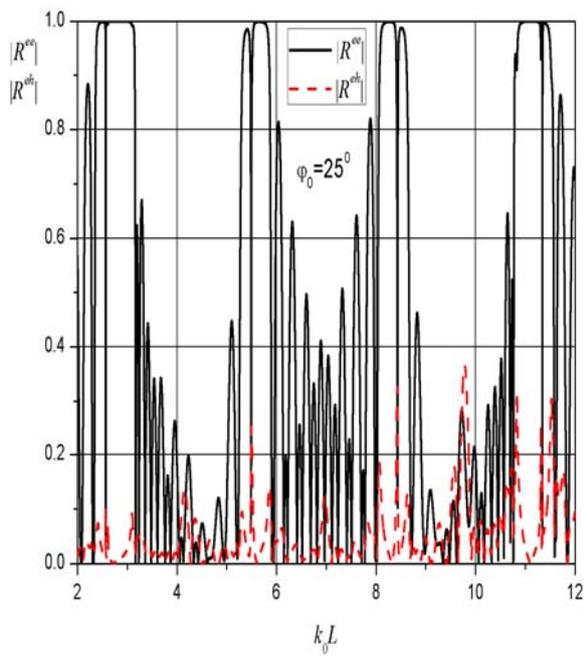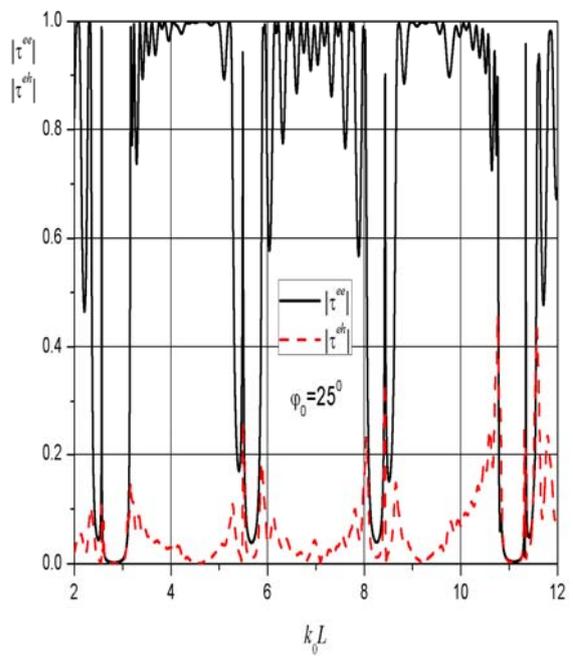

(c) (d)

Figure 6

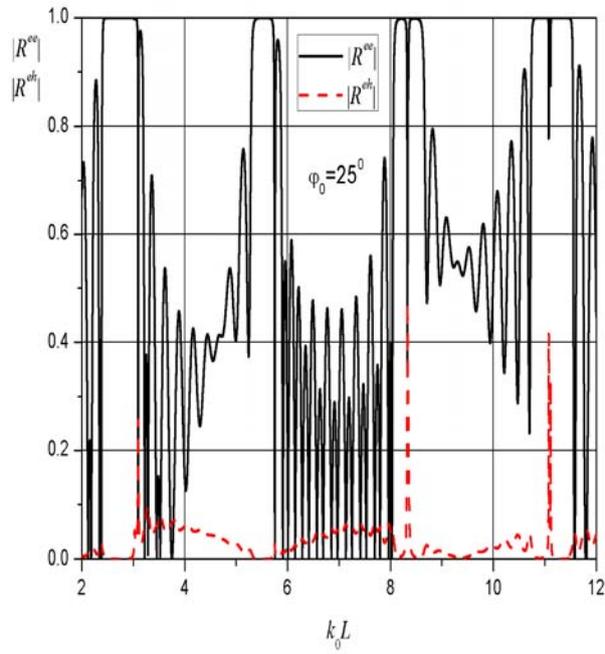

(a)

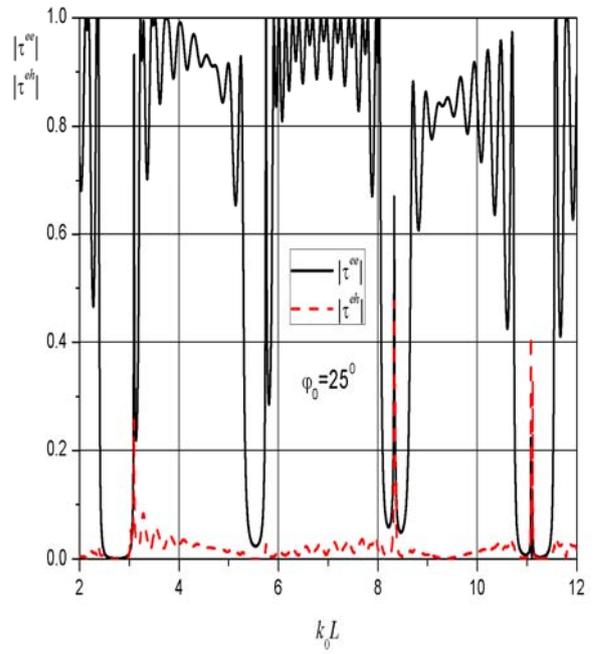

(b)

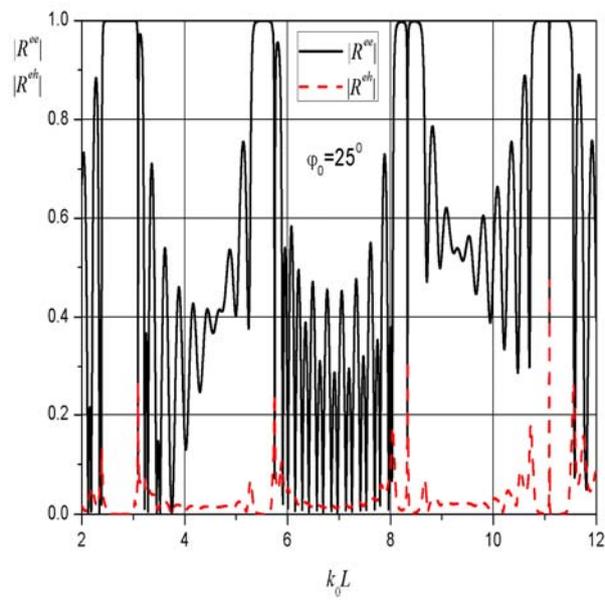

(c)

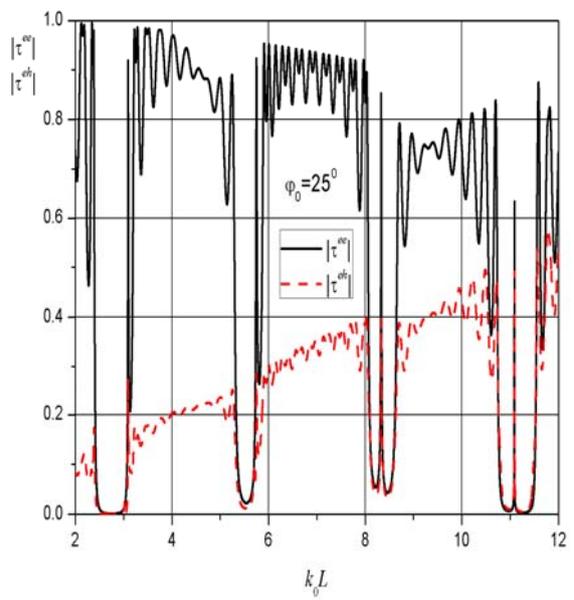

(d)

Figure 7